\documentclass[twocolumn,showpacs,preprintnumbers,amsmath,amssymb]{revtex4}
\usepackage{graphics}
\usepackage{epsfig}
\usepackage{color}

\begin{document}

\title{Out-of-equilibrium phase re-entrance(s) in long-range interacting systems}

\author{F. Staniscia}
\affiliation{Dipartimento di Fisica, Universit\`a di Trieste, Italy 
and Sincrotrone Trieste, S.S. 14 km 163.5, Basovizza (Ts), Italy }

\author{P.H. Chavanis}
\affiliation{
Universit\'e de Toulouse, UPS, Laboratoire de Physique Th\'eorique (IRSAMC),
F-31062 Toulouse, France and
CNRS, Laboratoire de Physique Th\'eorique (IRSAMC), F-31062 Toulouse, France. Marta 3, 50139 Firenze, Italy
}

\author{G. De Ninno}
\affiliation{
Physics Department, Nova Gorica University, Nova Gorica, Slovenia and
Sincrotrone Trieste, S.S. 14 km 163.5, Basovizza (Ts), Italy }

\author{D. Fanelli}
\affiliation{Dipartimento di Energetica, 
Via S. Marta 3, 50139 Florence, Italy
}

\date{\today}

\begin{abstract}
Systems with long-range interactions display a short-time relaxation
towards Quasi Stationary States (QSSs) whose lifetime increases with
system size. The application of Lynden-Bell's theory of ``violent
relaxation'' to the Hamiltonian Mean Field model leads to the
prediction of out-of-equilibrium first and second order phase
transitions between homogeneous (zero magnetization) and inhomogeneous
(non-zero magnetization) QSSs, as well as an interesting phenomenon of
phase re-entrances. We compare these theoretical predictions with
direct $N$-body numerical simulations. We confirm the existence of
phase re-entrance in the typical parameter range predicted from
Lynden-Bell's theory, but also show that the picture is more
complicated than initially thought. In particular, we exhibit the
existence of secondary re-entrant phases: we find un-magnetized states
in the theoretically magnetized region as well as persisting
magnetized states in the theoretically unmagnetized region.
\end{abstract}

\maketitle

\section{Introduction}

In statistical physics, phase re-entrance is a quite typical
phenomenon occurring in many physical systems, such as spin-glasses,
colloids and polymers, in which there is a competition between
different entropic terms \cite{cp,guven,han,sellitto,ekiz,rad}. A phase
re-entrance is normally associated with inverse melting, a
counterintuitive phenomenon in which isobaric addition of heat causes
a disordered (e.g., liquid) phase to crystallize, the reverse of the
usual situation. Phase re-entrance occurs when, providing additional
heating to the system, the latter undergoes a new transition, from the
ordered to the disordered phase. The phenomenon of phase re-entrance
has been widely studied at thermodynamic equilibrium in systems whose
constituents interact through short-range
forces. 

In this paper we give evidence to the existence of phase
re-entrance also in the case of long-range interacting systems in
out-of-equilibrium dynamical conditions.

Long-range interactions are such that the two-body interaction
potential decays at large distances with a power-law exponent which is
smaller than the space dimension. The dynamical and thermodynamical
properties of these systems were poorly understood until a few years
ago, and their study was essentially restricted to astrophysics (stellar systems) 
and two-dimensional turbulence (large-scale
vortices) \cite{csr}. Later, it was recognized that long-range systems
exhibit universal, albeit unconventional, equilibrium and
out-of-equilibrium features \cite{Dauxois_al02}. It is for instance
well known that such systems get trapped in long-lasting
Quasi-Stationary States (QSS)
\cite{henon,hc,ts,staquet,kn,Latora,Yama_04,campa}, before relaxing to thermal equilibrium. The
duration of a QSS increases with the number of particles $N$ in the
system. Remarkably, when the thermodynamic limit ($N
\rightarrow \infty$) is performed before the infinite time limit ($t
\rightarrow +\infty$), the
system remains permanently trapped in QSSs. As a consequence, QSSs
represent the only accessible experimental dynamical regimes for
systems composed by a large number of long-range interacting
particles. This includes systems of paramount importance, such as
non-neutral plasmas confined by a strong magnetic field \cite{hd,brands},
free-electron lasers \cite{Barre_pre_04} and ion particle beams
\cite{Benedetti}. The ubiquity of QSSs has originated an intense
debate \cite{assisi} about the mechanisms responsible for their
emergence, their persistence, and their eventual evolution towards
statistical equilibrium. In fact, QSSs keep memory of the initial
condition and, as a consequence, they cannot be interpreted by making
use of the classical Boltzmann-Gibbs approach. 

In a series of recent
papers \cite{Chavanis_EPJB_06,Antoniazzi_pre_07,prl_andreaduccio,Antoniazzi_prl_07,reentrant}, an approximate analytical theory
based on the Vlasov equation and inspired by a seminal work of
Lynden-Bell \cite{Lynden-Bell} in astrophysics has been proposed. This
is a fully predictive approach, enabling one to explain the emergence
and the properties of QSSs from first principles \cite{phcthesis}.

\begin{figure}
   \centering \includegraphics[width = 6cm,angle=-90]{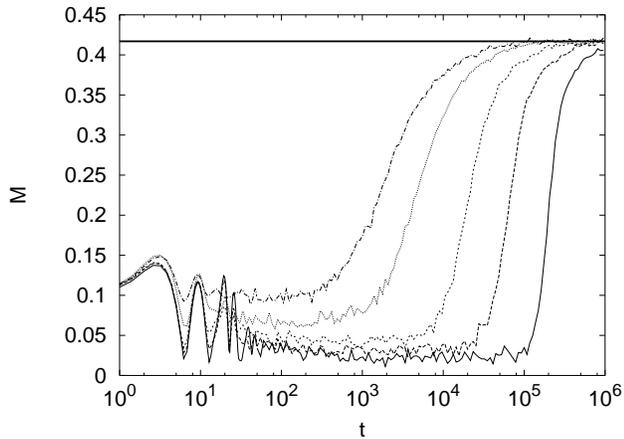}
   \caption{ Magnetization as a function of time, for systems with
   different sizes. The continuous, dashed, short-dashed, dotted and 
   dot-dashed lines correspond, respectively, to 
   $N = 2\cdot10^4, 10^4,5\cdot10^3,2\cdot10^3,10^3$. We see that QSSs start
   approximately at $t = 40$, and their duration increases with the
   the system size; eventually, they relax to Boltzmann equilibrium
   (thick line). Simulations are performed starting
   from a two-level distribution with energy
   $U=0.6400$, and averaging over different system realizations with the same initial distribution 
   ($50$, $100$, $200$, $500$ and $1000$, respectively). }  \label{fig:evo}
\end{figure}

In this paper we utilize a well-known hamiltonian toy model, the
so-called Hamiltonian Mean Field (HMF) model \cite{Antoni_Ruffo}, to
demonstrate that phase re-entrance(s) may also occur in a long-range
interacting system, dynamically trapped in a QSS. The HMF describes
the motion of $N$ rotators, coupled through an equal strength cosine
interaction. The Hamiltonian reads
\begin{equation}
\label{hamil}
H = \frac{1}{2} \sum_{j =1}^{N} p_{j}^2 + \frac{1}{2N}
\sum_{i,j=1}^{N} [ 1-\cos ( \theta_j - \theta_i)]
\end{equation}
where $\theta_j$ represents the orientation of the $j$-th rotator and $p_j$ stands for its conjugated momentum. To monitor the evolution
of the systems it is customary to introduce the magnetization, an order parameter defined as
\begin{equation}
M = \frac{\left|\sum_i \mathbf{m}_i\right|}{N} \quad \mbox{where}
\quad \mathbf{m}_i = ( \cos \theta_i , \sin \theta_i) \label{l}.
\end{equation}
The infinite-range coupling between rotators, provides the system with all typical characteristics of a long-range system, as clearly displayed in
Fig. \ref{fig:evo}. Here, the magnetization is monitored as a function of time: after an initial ``violent'' relaxation, the system reaches a QSS,
which is followed by a slow relaxation towards Boltzmann statistical equilibrium. The larger the system, the longer the intermediate phase where it remains confined before reaching the final equilibrium.

The paper is organized as follows. In Sec. \ref{sect2}, we present the
continuous Vlasov picture and review the maximum entropy principle
based on the Lynden-Bell approach. This theoretical setting is used to
obtain the HMF phase diagrams in different representations from which
out-of-equilibrium phase transitions
\cite{Chavanis_EPJB_06,Antoniazzi_pre_07,prl_andreaduccio,Antoniazzi_prl_07,reentrant}
and phase re-entrance
\cite{Chavanis_EPJB_06,reentrant} can be predicted.  In
Sec. \ref{sect3}, these theoretical predictions are compared to
$N$-body simulations based on
\eqref{hamil}. Finally, in Sec. \ref{sect4} we sum up our results
and draw our conclusions.

\section{Out-of-equilibrium phase re-entrance: the prediction of Lynden-Bell theory}
\label{sect2}

\subsection{General theory and two-levels approximation}

In a recent series of papers
\cite{Chavanis_EPJB_06,Antoniazzi_pre_07,prl_andreaduccio,Antoniazzi_prl_07,reentrant},
an approximate analytical theory based on the Vlasov equation has been
proposed for the HMF model stemming from the seminal work of
Lynden-Bell
\cite{Lynden-Bell}.  This
is a fully predictive approach, justified from first principles, which
captures most of the peculiar traits of the HMF out-of-equilibrium
dynamics.  The philosophy of the proposed approach is reviewed in the
following.

In the limit of $N \rightarrow \infty$, the HMF dynamics can be formally described using the Vlasov equation
\begin{equation}
\frac{\partial f}{\partial t} + p\frac{\partial f}
{\partial \theta} - \left( M_x[f] \sin \theta - M_y[f]\cos \theta \right) \frac{\partial f}{\partial p}=0,
\label{eq:VlasovHMF}
\end{equation}
where $f(\theta,p,t)$ is the one-body microscopic distribution function (DF), and the two components of the  magnetization
are respectively given by
\begin{eqnarray}
\label{magnetization}
M_x[f]&=&\int f \cos \theta d\theta dp, \\
M_y[f]&=&\int f \sin \theta d\theta dp. \nonumber
\end{eqnarray}
The mean field energy can be expressed as
\begin{eqnarray}
\label{energy}
U=\frac{1}{2}\int f p^{2}d\theta dp-\frac{M_x^2+M_y^2}{2}+\frac{1}{2}.
\end{eqnarray}
Working in this setting, it can be then hypothesized that QSSs
correspond to stable stationary equilibria of the Vlasov equation on a
coarse-grained scale. Lynden-Bell's idea goes as follows. The Vlasov
dynamics induces a progressive filamentation of the initial single
particle distribution profile, i.e. the continuous counterpart of the
discrete $N$-body distribution, which proceeds at smaller and smaller
scales without reaching an equilibrium. Conversely, at a coarse-grained
level, the process comes to an end, and the distribution function
$\bar{f}_{QSS}(\theta,p,t)$, averaged over a finite grid, eventually
converges to an asymptotic form. Following Lynden-Bell, one can then
associate a mixing entropy to this process. Assuming ergodicity (i.e.,
efficient mixing), $\bar{f}_{QSS}(\theta,p)$, is obtained by
maximizing the mixing entropy, while imposing the conservation of
Vlasov dynamical invariants. It is worth emphasizing that the
prediction of the QSS depends on the details of the initial condition
\cite{Chavanis_06}, not only on the values of energy and mass as for the
Boltzmann statistical equilibrium state. This is due to the fact that
the Vlasov equation admits an infinite number of invariants, i.e. the
Casimirs or, equivalently, the moments ${\cal M}_{n}=\int
\overline{f^{n}}d\theta dp$ of the fine-grained
distribution function. The first moment ${\cal M}_{1}={\cal M}$ is
just the normalization of the distribution function, and we can refer
to it as the conservation of the total mass.

In the case of a two-levels initial condition, where the fine-grained
DF takes only two values $f=f_0$ and $f=0$, the invariants reduce to
${\cal M}$ and $U$, since the moments ${\cal M}_{n>1}$ can all be
expressed in terms of ${\cal M}$ and $f_{0}$ as ${\cal M}_{n}=\int
\overline{f^{n}}d\theta dp=\int \overline{f_{0}^{n-1} \times f}d\theta
dp=\int f_0^{n-1} \overline{f} d\theta dp=f_0^{n-1} {\cal M}$. For the
specific case at hand, the Lynden-Bell entropy is then explicitly
constructed from the coarse-grained distribution function $\bar{f}$
and reads \cite{Chavanis_EPJB_06,Chavanis_pro}:
\begin{equation}
S[\bar{f}]=-\int \!\!{\mathrm d}p{\mathrm d}\theta \,
\left[\frac{\bar{f}}{f_0} \ln \frac{\bar{f}}{f_0}
+\left(1-\frac{\bar{f}}{f_0}\right)\ln
\left(1-\frac{\bar{f}}{f_0}\right)\right].
\label{eq:entropieVlasov}
\end{equation}
We thus have to solve the maximization problem
\begin{eqnarray}
\label{vr7}
\max_{\overline{f}}\lbrace S[\overline{f}]\quad |\quad U[\overline{f}]=U,
{\cal M} [\overline{f}]=1 \rbrace.
\end{eqnarray}
This maximization problem assures that the distribution function is
thermodynamically stable (most probable macrostate) in the sense of
Lynden-Bell \cite{Lynden-Bell}. The maximization problem (\ref{vr7}),
for an arbitrary functional of the form $S[f]=-\int C(f)\, d\theta dp$
where $C$ is convex, also forms a criterion of formal nonlinear dynamical
stability with respect to the Vlasov equation \cite{Chavanis_EPJB_05}.
In the two levels approximation, where the Lynden-Bell entropy is a
functional of $\overline{f}$, the criteria of dynamical and
Lynden-Bell thermodynamical stability coincide. From
Eq. (\ref{eq:entropieVlasov}), we write the first order variations as
\begin{eqnarray}
\label{vr8}
\delta S-\beta\delta U-\alpha\delta {\cal M}  =0,
\end{eqnarray}
where the inverse temperature $\beta=1/T$ and the ``chemical
potential'' $\alpha$ are Lagrange multipliers associated to the
conservation of energy and mass. Requiring that this functional is
stationary, one obtains the following distribution
\cite{Chavanis_EPJB_06,Antoniazzi_prl_07}:
\begin{multline}
  \label{eq:barf}
  \bar{f}_{\text{QSS}}(\theta,p)=  \\
  \frac{f_0}{ 1+e^{\displaystyle\beta (p^2/2 -M_x[\bar{f}_{\text{QSS}}]\cos\theta-M_y[\bar{f}_{\text{QSS}}]\sin\theta)+\alpha}}.
\end{multline}
As a general remark, it should be emphasized the ``fermionic'' form of
the distribution, which arises because of the form of the
entropy. Notice also that the magnetization is related
self-consistently to the distribution function by
Eq. (\ref{magnetization}), and the problem hence amounts to solving an
{\it integro-differential} equation. In doing so, we have also to make
sure that the critical point corresponds to an entropy maximum, not to
a minimum or a saddle point. Let us now insert expression
\eqref{eq:barf} into the energy and normalization constraints and use
the definition of magnetization \eqref{magnetization}. Further,
defining $\lambda=e^{\alpha}$ and ${\mathbf m}=(\cos \theta, \sin
\theta)$ yields 
\footnote{These parameters are related to those introduced
in \cite{Chavanis_EPJB_06} by $U=\epsilon/4+1/2$, $\beta=2\eta$,
$f_0=\eta_{0}/N=\mu/(2\pi)$, $k=2\pi/N$, $x=\Delta\theta$,
$y=(2/\pi)\Delta p$ and the functions $F$ in \cite{Antoniazzi_prl_07} are
related to the Fermi integrals by
$F_{k}(1/y)=2^{(k+1)/2}y I_{(k-1)/2}(y)$.}:
\begin{eqnarray}
\label{eq:cond0}
&f_0& \sqrt{\frac{{2}}{{\beta}}}  \int {\mathrm d} \theta  I_{-1/2}\left(\lambda  e^{-\beta {\mathbf M} \cdot {\mathbf
m}}\right) = 1, \\
\label{eq:cond3}
&f_0& \frac{1}{2} \left (\frac{2}{\beta}\right )^{3/2} \int {\mathrm d} \theta
I_{1/2}\left(\lambda e^{-\beta {\mathbf M} \cdot {\mathbf m}}\right)
= U+\frac{M^2 - 1}{2}, \nonumber\\
\label{eq:cond1}
&f_0& \sqrt{\frac{{2}}{{\beta}}}   \int {\mathrm d} \theta \cos \theta
I_{-1/2}\left(\lambda e^{-\beta {\mathbf M} \cdot {\mathbf m}}\right)
= M_x, \nonumber \\
\label{eq:cond2}
&f_0& \sqrt{\frac{{2}}{{\beta}}} \int {\mathrm d} \theta \sin \theta
I_{-1/2}\left(\lambda e^{-\beta {\mathbf M} \cdot {\mathbf m}}\right)
= M_y, \nonumber
\end{eqnarray}
where we have defined the Fermi integrals
\begin{eqnarray}
\label{es5}
I_{n}(t)=\int_{0}^{+\infty}\frac{x^{n}}{1+t e^{x}}dx.
\end{eqnarray}
Its asymptotic limits are recalled in \cite{Chavanis_EPJB_06}.

If we consider spatially homogeneous configurations ($M_{QSS}=0$), the Lynden-Bell distribution becomes
\begin{eqnarray}
  \label{eq:barfhom}
  \bar{f}_{\text{QSS}}(p)=
  \frac{f_0}{ 1+\lambda e^{\beta p^2/2}}.
\end{eqnarray}
In the non degenerate limit $\lambda\rightarrow +\infty$, the latter reduces to the Boltzmann distribution $\bar{f}=(\beta/2\pi)^{1/2}e^{-\beta p^2/2}$ and in the completely degenerate limit $\lambda\rightarrow 0$, it becomes a step function: $\bar{f}=f_0$ for $p<1/(4\pi f_0)$ and $\bar{f}=0$ otherwise. If we make use of Eqs. (\ref{eq:cond0}), we get the following parametric
relation, with parameter $\lambda$, between the inverse temperature $\beta$ and the energy $U$ (for a fixed value of $f_0$) \cite{Chavanis_EPJB_06}:
\begin{eqnarray}
\label{G1}
\left ( U - \frac{1}{2}\right) 8\pi^{2}f_0^2=  \frac{I_{1/2}(\lambda)}{I_{-1/2}(\lambda)^3} \equiv G(\lambda) ,\\
\frac{\beta}{8 \pi^2 f_0^2}=  I_{-1/2}(\lambda)^2, \nonumber
\end{eqnarray}
where $G(\lambda)$ is a universal function monotonically increasing with $\lambda$. A solution of the above equation certainly exists provided that
\begin{equation}
\label{G2}
\left (U-\frac{1}{2} \right ) 8 \pi^2 f_0^2 \ge G(0).
\end{equation}
To compute $G(0)$ one can use the asymptotic expansions of the Fermi
integrals. This yields $G(0)=1/12$. Therefore, the homogeneous
Lynden-Bell distribution with fixed $f_0$ exists only for
\cite{Chavanis_EPJB_06}:
\begin{eqnarray}
\label{miny}
U\ge U_{min}(f_0)\equiv\frac{1}{96\pi^2 f_0^{2}}+\frac{1}{2}.
\end{eqnarray}
%For a fixed value of $f_0$, the completely degenerate limit ($\lambda=0$) corresponds to
%$U=U_{min}(f_0)$ and $\beta\rightarrow +\infty$.  The non degenerate
%limit ($\lambda\rightarrow +\infty$) corresponds to $U\rightarrow
%+\infty$ and $\beta\rightarrow 0$. In this limit, the caloric curve
%(\ref{G1}) takes the classical form $U\simeq 1/(2\beta)+1/2$
%\cite{Chavanis_EPJB_06,reentrant}.

The completely degenerate limit ($\lambda=0$) of the homogeneous distribution corresponds to:

(i) $U=U_{min}(f_0)$ and $\beta\rightarrow +\infty$ for any $f_0$.

The non degenerate limit ($\lambda\rightarrow +\infty$) of the homogeneous distribution corresponds to:

(i)  $U\rightarrow
+\infty$ and $\beta\rightarrow 0$  for any $f_0$. In this limit, the caloric curve
(\ref{G1}) takes the classical form $U\simeq 1/(2\beta)+1/2$
\cite{Chavanis_EPJB_06,reentrant}. 

(ii) $f_0\rightarrow +\infty$ for any $U$.

Let us now address the problem of stability of the homogeneous Lynden-Bell distribution. In  \cite{Chavanis_EPJB_06}, it has been shown that the homogeneous Lynden-Bell distribution is stable if, and only if:
\begin{eqnarray}
\label{vs7}
I_{-1/2}(\lambda)\lambda |I_{-1/2}'(\lambda)|\le \frac{1}{(2\pi f_0)^2}.
\end{eqnarray}
If the distribution function satisfies (\ref{vs7}) then it is both
Lynden-Bell thermodynamically stable (entropy maximum) and
formally non-linearly dynamically stable
\cite{Chavanis_EPJB_05}. Otherwise, it is Lynden-Bell
thermodynamically unstable (saddle point of entropy) and linearly
dynamically unstable. For a given $f_0$, the relation (\ref{vs7})
determines the range of $\lambda$ for which the homogeneous
distribution is stable/unstable. Then, using Eqs. (\ref{G1}), we can
determine the range of corresponding energies. Specifically, the
critical curve $U_{c}(f_0)$ separating stable and unstable homogeneous
Lynden-Bell distributions is given by the parametric equations
\cite{Chavanis_EPJB_06}:
\begin{eqnarray}
\label{vs7b}
I_{-1/2}(\lambda)\lambda |I_{-1/2}'(\lambda)|= \frac{1}{(2\pi f_0)^2}, \\
U-\frac{1}{2}= \frac{1}{8 \pi^2 f_0^2} \frac{I_{1/2}(\lambda)}{I_{-1/2}(\lambda)^{3}}, \nonumber
\end{eqnarray}
where $\lambda$ goes from $0$ (completely degenerate) to $+\infty$
(non degenerate). In fact, the criterion (\ref{vs7}) only proves that
$f$ is a {\it local} entropy maximum at fixed mass and energy. If
several local entropy maxima are found (for example, homogeneous and
inhomogeneous Lynden-Bell distributions), we must compare their
entropies to determine the stable state (global entropy maximum) and
the metastable states (secondary entropy maxima). For systems with
long-range interactions, metastable states have in general very long
lifetimes, scaling like $e^{N}$, so that they are stable in practice
and must absolutely be taken into account
\cite{Antoni_04, Chavanis_AA_05}. For this reason, (out-of-equilibrium) stability diagrams do not coincide with phase diagrams. In fact, the latter require a careful investigation of metastable states.

\subsection{Phase diagram in the $(f_0,U)$ plane}
\label{fu}

\begin{figure}
%[htbp]
\centering
%\vspace*{3em}
\includegraphics[width=5.5cm,angle=-90]{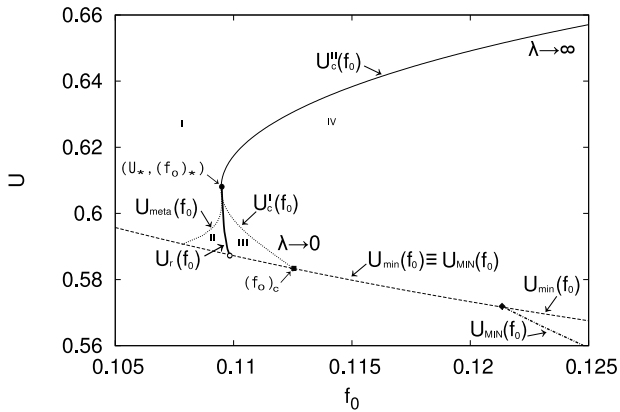}
\caption{Phase diagram in the $(f_0,U)$ plane. The homogeneous phase only exists above the line $U_{min}(f_0)$. The stability curve $U_{c}(f_0)$ is parameterized by $\lambda$. For $\lambda\rightarrow 0$ (completely degenerate limit), we get $f_0=(f_0)_c=1/(2\pi\sqrt{2})$ and $U_{c}=7/12$. For  $\lambda\rightarrow +\infty$ (non degenerate limit), we get $f_0\rightarrow +\infty$ and $U_{c}=3/4$. On the left of this curve, the homogeneous phase is stable and on the right of this curve it is unstable. The stability curve is divided in two parts, i.e. $U^{\prime}_c(f_0)$ and $U^{\prime \prime}_c(f_0)$, by the
tricritical point (full round dot) located at $((f_0)_*,U_*)$. The
continuous line corresponds to the second-order transition line while
the dotted lines correspond to the borders of the metastable
region. The thick line represents the first-order transition line. All
these lines divide the diagram in four regions.  In region (I), the
homogeneous phase is stable and the inhomogeneous phase does not
exist; in (II), the homogeneous phase is stable and the inhomogeneous
phase metastable; in (III), the homogeneous phase is metastable and
the inhomogeneous phase stable; in (IV) the homogeneous phase is
unstable and the inhomogeneous phase stable.  $U_{min}(f_0)$ is the
line below which the homogeneous phase does not exist, and
$U_{MIN}(f_0)$ is the lowest accessible value of energy for a
rectangular water bag initial condition (see Sec. \ref{min}). The
square dot is $((f_0)_c,U_c((f_0)_c))$, the diamond is $((f_{0})_{m},U_m)$
(Sec. \ref{min}), and the empty round dot is
$((f_{0})_{r},U_r((f_{0})_{r}))$. For $f_0 \in
\lbrack (f_{0})_{*},(f_{0})_{c}\rbrack $ there is a re-entrant phase.}
%Stability diagram of the homogeneous phase in the $(f_0,U)$ plane. The
%homogeneous phase only exists above the line $U_{min}(f_0)$. It is stable above the line
%$U_{c}(f_0)$ and unstable below it. [REVIEW THE FOLLOWING AFTER HAVING PLOTTED METASTABLE REGIONS: For fixed $f_{0}\in \lbrack (f_{0})_{*},(f_{0})_{c}\rbrack$,
%there is a re-entrant phase as energy is progressively decreased: the homogeneous phase is stable
%for $U>U_{c}^{(2)}(f_0)$, unstable for
%$U_{c}^{(2)}(f_0)>U>U_{c}^{(1)}(f_0)$ and {\it stable again} for
%$U_{c}^{(1)}(f_0)>U\ge U_{min}(f_0)$. ]}
\label{muepsilon}
\end{figure}

The phase diagram of the Lynden-Bell distribution in the $(f_0,U)$ plane is shown in Fig.
\ref{muepsilon}. We have also plotted the stability curve
$U_{c}(f_0)$ of the homogeneous phase (split in two parts, $U^{\prime}_c(f_0)$ and $U^{\prime \prime}_c(f_0)$) defined by Eqs. (\ref{vs7b}) and
parameterized by $\lambda$. On the left of this curve, the homogeneous
phase is stable (maximum entropy state) and on the right of this curve
it is unstable (saddle point of entropy)
\footnote{Here, the term ``unstable'' means that the homogeneous
Lynden-Bell distribution is not a maximum entropy state, i.e. (i) it
is not the most mixed state, and (ii) it is dynamically unstable with
respect to the Vlasov equation. Hence, it should not be reached as a
result of violent relaxation. One possibility is that the system
converges to the spatially inhomogeneous Lynden-Bell distribution
\eqref{eq:barf} with $M_{QSS}\neq 0$ which is the maximum entropy
state (most mixed) in that case. Another possibility, always to
consider, is that the system does not converge towards the maximum
entropy state, i.e. the relaxation is incomplete
\cite{Chavanis_EPJB_06}.}. For $f_0\rightarrow +\infty$, we are in
the non-degenerate limit $\lambda\rightarrow +\infty$ and the
stability criterion (\ref{vs7}) is equivalent to $\lambda^2\ge 4\pi^3
f_0^2$. Using Eq. (\ref{G1}), this yields $U\ge U_{c}=3/4$. This is
the critical energy associated to the Maxwell distribution. On the
line of minimum energy $U=U_{min}(f_0)$, we are in the completely
degenerate limit ($\lambda\rightarrow 0$) and the stability criterion
(\ref{vs7}) is equivalent to $f_0\le (f_0)_c=1/(2\pi\sqrt{2})$. Using
Eq. (\ref{G1}), this yields $U\ge U_{c}=7/12$. This is the critical
energy associated to the spatially homogeneous water-bag distribution
(step function). Therefore, the minimum energy curve $U_{min}(f_0)$
crosses the stability curve $U_{c}(f_0)$ at $((f_0)_c,U_{c}) \simeq
(0.1125, 0.5833)$.

If we now take into account Lynden-Bell's inhomogeneous states,
solving numerically Eqs. (\ref{eq:cond0}), we find that the phase
diagram displays first-order and second-order phase transitions.  The
curve $U_{c}(f_0)$ splits in two curves $U^{\prime}_c(f_0)$ and
$U^{\prime \prime}_c(f_0)$. In the case of a second-order phase
transition, the stability threshold corresponds to the transition
between a homogeneous and an inhomogeneous distribution. The
second-order phase transition corresponds to the branch
$U''_c(f_0)$. On the other hand, for a first-order phase transition,
as we have the coexistence of two entropy maxima, the stability
condition of the homogenous phase is no more sufficient to find the
transition line, which has to be calculated by making a comparison
between the two entropy maxima. This procedure has been followed to
plot the line $U_{r}(f_0)$ in Fig. \ref{muepsilon}. This line
of first-order phase transition is reached when the  homogeneous and
inhomogeneous phases have the same entropy. The line $U_{r}(f_0)$
(first order) and the line $U_{c}^{\prime \prime}(f_0)$ (second order) merge together
at a tricritical point, located at
$((f_{0})_*,U_*)\simeq(0.10947,0.608)$ and corresponding to $\lambda_*=0.024$. 
We have also plotted the curves $U_{c}'(f_0)$ and $U_{meta}(f_0)$  giving the lateral edges of the
metastability regions for the homogeneous and inhomogeneous phases
(see figure caption for more details).

In conclusion, the second order phase transition occurs for a range of
values of $U(f_0)$ bounded by the tricritical point $(U_*,(f_0)_*)$,
and by $U_c = 3/4$, reached for $f_0 \rightarrow +\infty$. For
$U>U_c=3/4$, the Lynden-Bell theory always predicts a homogeneous
phase (for any value of $f_0$). For $f_0<(f_0)_*$, the homogeneous
phase is always stable (for any $U\ge U_{min}(f_0)$).  For
$f_0>(f_0)_c$, the homogeneous phase is unstable for $U_{min}(f_0)\le
U< U_c(f_0)$ and stable for $ U> U_c(f_0)$. The first order phase
transition occurs for a range of $U_c(f_0)$ bounded by the tricritical
point $(U_*,(f_0)_*)$ and by the point $((f_{0})_{r},U_r
((f_{0})_{r}))\simeq(0.1098,0.5875)$. As can be seen in
Fig. \ref{muepsilon}, the theory predicts a phase re-entrance, for a
set of values of $f_0$ $\in \lbrack (f_{0})_{*},(f_{0})_{c}\rbrack
$. This means that, increasing $U$ in the diagram at fixed $f_0$ $\in
\lbrack (f_{0})_{*},(f_{0})_{c}\rbrack $, the homogeneous phase is
(meta)stable for $U_{min}(f_0) < U < U_c'(f_0)$, unstable for $
U_c'(f_0) < U <U^{\prime \prime}_c(f_0)$, and stable again for $ U >
U^{\prime
\prime}_c(f_0)$. Note that in the metastability
region $U_{min}(f_0) < U < U_c'(f_0)$, the system can be found either
in the homogeneous or inhomogeneous phase depending on how it has been
prepared initially (recall that metastable states are highly robust
for systems with long-range interactions). By contrast, for $
U_c'(f_0) < U <U^{\prime \prime}_c(f_0)$, the theory predicts an
inhomogeneous phase and for $U > U^{\prime \prime}_c(f_0)$ a
homogeneous phase.

%For $U>3/4$, the homogeneous phase is always stable
%(maximum entropy state) for any $f_0$ (in fact, the curve $U_{c}(f_{0})\rightarrow 3/4$ for $f_0\rightarrow +\infty$).
%For $f_0<(f_0)_{*}$, the homogeneous phase is always stable (maximum entropy state), for any $U\ge U_{min}(f_0)$.
%For $f_0>(f_0)_{c}$ (where $(f_0)_{c}$ is the value of $f_0$ at the intersection between $U_{c}(f_{0})$ and $U_{min}(f_0)$), the
%homogeneous phase is stable for $U>U_{c}(f_0)$ and unstable for $U_{min}(f_0)\le U<U_{c}(f_0)$. This range of parameters corresponds
%to the second order phase transition. For $(f_0)_{*}<f_0<(f_0)_{c}$, the system displays a re-entrant stability regime:
%the homogeneous phase is stable for $U>U_{c}^{(2)}(f_0)$, unstable for $U_{c}^{(2)}(f_0)>U>U_{c}^{(1)}(f_0)$ and {\it stable again}
%for $U_{c}^{(1)}(f_0)>U\ge U_{min}(f_0)$.
%In the diagram is plotted also the curve of the minimum accessible energy for the homogeneous phase $U_{min}(f_0)$ and the metastable regions.
%These are separated by the fist order transition line, which meets the last line at ($(f_{0})_{r} = ... $, $(f_{0})_{r}= ...$), and the
%second order transition line at the tricritical point (located at $(f_{0})_{*}=0.10947..., U_{*}=0.608...$).
%This last line, because of the order of the transition, coincide with the stablity line.
%This means that Lynden-Bell theory predicts a ``phase'' re-entrance for a set of values of $f_0$, bounded by $(f_{0})_{*}$ and $(f_{0})_{r}$.
%This diagram is fully consistent with the ($U$,$M_0$) one.
%\cite{reentrant}

\subsection{Phase diagram in the $(M_0,U)$ plane}
\label{mu}

The preceding results are valid for {\it any} initial condition with
two phase levels $f=0$ and $f=f_0$, whatever the number of patches and
their shape. In the two-levels case, the relevant control parameters
are $(f_0,U)$ \cite{Chavanis_EPJB_06}. They fully specify the
Lynden-Bell equilibrium state from the initial condition. This means
that, assuming ergodicity, the system remembers the initial condition
through the values of these parameters. In this sense, the general
phase diagram in the two-levels case is the one represented in
Fig. \ref{muepsilon}.

Now, many numerical simulations of the $N$-body system
\cite{prl_andreaduccio}, or of the Vlasov equation
\cite{Antoniazzi_prl_07}, have been performed starting from a family
of rectangular water-bag distributions. The latter correspond to
assuming a constant value $f_{0}$ inside the phase-space domain $D$:
\begin{equation}
D = \{(\theta , p ) \in [ -\pi , \pi] \times [- \infty, + \infty] \mid
|\theta| < \Delta \theta, |p| < \Delta p \}\label{g}
\end{equation}
and zero outside. Here $ 0 \le \Delta \theta \le \pi$ and $ \Delta p
\ge 0 $.  The normalization condition results in
\begin{equation}
\label{f0d}
f_0 = \frac{1}{4 \Delta
\theta \Delta p}.
\end{equation}
Notice that, for this specific choice, the initial magnetization
$M_{0}$ and the energy density $U$ can be expressed as functions of
$\Delta\theta$ and $\Delta p$ as
\begin{equation}
U=\frac{(\Delta p)^2}{6}+\frac{1-(M_0)^2}{2}\label{i} \: ,
\end{equation}
\begin{equation}
 M_0=\frac{\sin(\Delta \theta)}{\Delta \theta}\label{h} \; .
\end{equation}
For the case under scrutiny, $0 \leq M_{0} \leq 1$ and $U \geq
U_{MIN}(M_0)\equiv (1 - M_0^2)/2$. The energy $U_{MIN}(M_0)$
represents the absolute minimum accessible energy for a rectangular
water-bag distribution with magnetization $M_{0}$. The initial
configuration is completely specified by the variables
$(\Delta\theta,\Delta p)$ or, equivalently, by the variables
$(M_0,U)$. On the other hand, for the determination of the Lynden-Bell
equilibrium state, only the variables $(f_0,U)$ matter. Now, we note
that different values of $(M_0,U)$ can correspond to the same
$(f_0,U)$ and, consequently, to the {\it same} Lynden-Bell
equilibrium (see Sec. \ref{connection}). Therefore, the use of these
variables leads to some redundances. Nevertheless, their advantage is
that they are more directly related to physically accessible
parameters. In any case, it is of interest to compare the two phase
diagrams in $(f_0,U)$ and $(M_0,U)$ planes to see their similarities
and differences.

For the rectangular water-bag initial condition, using
Eqs. (\ref{f0d}) and (\ref{i}), we can express $f_0$ as a function of
$M_0$ and $U$ by
\begin{eqnarray}
\label{f0mu}
f_0^2=\frac{1}{48\lbrack (2U-1)(\Delta\theta)^2+\sin^{2}\Delta\theta\rbrack},
\end{eqnarray}
where $\Delta\theta$ is related to $M_{0}$ by
Eq. (\ref{h}). Inserting this expression in Eqs. (\ref{G1}), we
obtain after some algebra the caloric curve $T(U)$ for fixed $M_0$
parameterized by $\lambda$:
\begin{eqnarray}
\label{calb}
&& U-\frac{1}{2}=\frac{\sin^{2}\Delta\theta}{\frac{\pi^{2}}{6}\frac{I_{-1/2}
(\lambda)^3}{I_{1/2}(\lambda)}-2(\Delta\theta)^2}, \nonumber \\
&& \beta=\frac{1}
{\sin^{2}\Delta\theta}\left (\frac{\pi^{2}}{6}I_{-1/2}
(\lambda)^2-2(\Delta\theta)^{2}\frac{I_{1/2}(\lambda)}{I_{-1/2}(\lambda)}\right ).\nonumber \\
\end{eqnarray}
Using Eqs.  (\ref{miny}) and (\ref{f0mu}), we find that the homogeneous Lynden-Bell distribution with fixed $M_0$ exists if and only if
\begin{eqnarray}
\label{cn5}
U\ge U_{min}(M_0)\equiv \frac{1}{2}\left (\frac{\sin^{2}\Delta\theta}{\pi^{2}-(\Delta\theta)^{2}}+1\right ).
\end{eqnarray}
This result can also be obtained from Eq. (\ref{calb}) by considering the limit $\lambda\rightarrow 0$.

The completely degenerate limit ($\lambda=0$) of the homogeneous distribution corresponds to: 

(i) $U=U_{min}(M_0)$ and $\beta\rightarrow +\infty$ for any $M_0$. 

(ii) $M_0=0$ for any $U$.

The non degenerate limit ($\lambda\rightarrow +\infty$) of the homogeneous distribution corresponds to: 

(i)  $U\rightarrow
+\infty$ and $\beta\rightarrow 0$  for any $M_0$. In that case $U\simeq 1/(2\beta)+1/2$.

(ii) $M_0=1$ for any $U$. In that case $f_0\rightarrow +\infty$.

On the other hand, regrouping all the preceding results, the critical
curve $U_{c}(M_0)$ separating stable and unstable homogeneous
Lynden-Bell distributions is given by the parametric equations
\begin{eqnarray}
\label{cu1}
I_{-1/2}(\lambda)\lambda |I_{-1/2}'(\lambda)|= \frac{1}{(2\pi f_0)^2},
\end{eqnarray}
\begin{eqnarray}
\label{cu2}
 U-\frac{1}{2}=
\frac{1}{8 \pi^2 f_0^2} \frac{I_{1/2}(\lambda)}{I_{-1/2}(\lambda)^{3}},
\end{eqnarray}
\begin{eqnarray}
 \label{cu3}
f_0^2=\frac{1}{48\lbrack (2U-1)(\Delta\theta)^2+\sin^{2}\Delta\theta\rbrack},
\end{eqnarray}
\begin{eqnarray}
 \label{cu4}
M_{0} = \dfrac{\sin(\Delta\theta)}{\Delta\theta} ,
\end{eqnarray}
where $\lambda$ goes from $0$ (completely degenerate) to $+\infty$
(non degenerate).

\begin{figure}
%[htbp]
%  \centering
%\vspace*{3em}
\includegraphics[width=8cm,angle=0]{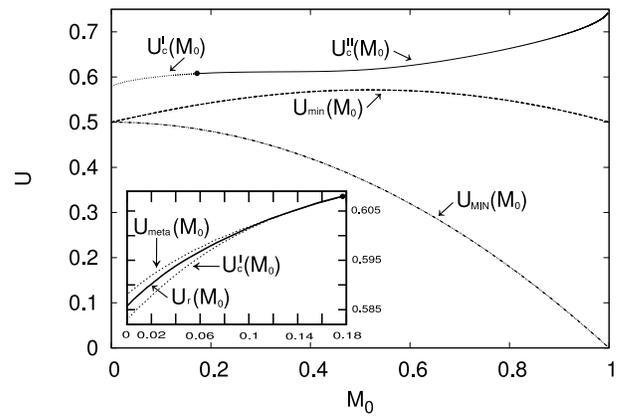}
  \caption{Phase diagram in the control parameter plane
  $(M_{0},U)$ for a rectangular water-bag initial
  profile. $U_{MIN}(M_0)$ is the absolute minimum energy, and the
  homogeneous phase only exists above the line $U_{min}(M_0)$. The
  stability curve $U_{c}(M_0)$ is parameterized by $\lambda$. For
  $\lambda\rightarrow 0$ (completely degenerate limit), we get $M_0=0$
  and $U_{c}=7/12$. For $\lambda\rightarrow +\infty$ (non degenerate
  limit), we get $M_0=1$ and $U_{c}=3/4$. Above this curve, the
  homogeneous phase is stable and below this curve it is unstable. The
  full dot is the tricritical point. In the inset is showed the region
  of the first order phase transition, indicated by the line
  $U_r(M_0)$, connected to the second order phase transition line
  $U_c''(M_0)$ by the tricritical point. The
  dotted lines represent the borders of the metastability
  region.}
  \label{fig:phase-diagram}
\end{figure}

The phase  diagram of the  Lynden-Bell distribution in
the $(M_0,U)$ plane is represented in Fig. \ref{fig:phase-diagram}. We
have first plotted the minimum accessible energy of the homogeneous
phase $U_{min}(M_0)$ defined by Eq. (\ref{cn5}). We have also plotted
the stability curve $U_{c}(M_0)$ of the homogeneous phase defined by
Eqs. (\ref{cu1})-(\ref{cu4}) and parameterized by $\lambda$. Above this curve the homogeneous
($M_{QSS}=0$) phase is stable (maximum entropy state) and below this
curve it is unstable (saddle point of entropy). For $M_0=1$, we are in
the non degenerate limit $\lambda \rightarrow +\infty$  (because $f_0\rightarrow +\infty$) and the
critical energy is $U_c=3/4$ (Maxwell distribution). For $M_0=0$, we are in the completely
degenerate limit $\lambda=0$ and the critical energy is $U_c=7/12$ (spatially homogeneous water bag).

If we now take into account Lynden-Bell's inhomogeneous states,
solving numerically Eqs. (\ref{eq:cond0}), we find that the phase
diagram displays first-order and second-order phase transitions.  The
second order phase transition corresponds to the branch $U''_c(M_0)$
and the first order phase transition to the branch $U_{r}(M_0)$.
These two lines merge together at the tricritical point, 
located at $((M_{0})_{*},U_*)\simeq(0.1757,0.608)$ corresponding to
$\Delta\theta_{*}=2.656...$ Using Eq. (\ref{f0mu}), we readily check
that this tricritical point $((M_{0})_{*},U_*)$ corresponds to the tricritical point
$((f_{0})_{*},U_*)$ in the $(f_0,U)$ plane. We have
also plotted in the inset the lateral edges $U'_{c}(M_0)$ and
$U_{meta}(M_0)$ of the metastability region associated to the first order
phase transition.

Therefore, the phase diagrams in $(f_{0},U)$ and $(M_{0},U)$ planes
are fully consistent (see Sec. \ref{connection} for more details) and
both display first and second phase transitions. The correctness of
the above analysis is assessed in \cite{prl_andreaduccio} where
numerical simulations are performed for different values of the system
size $N$. The transitions predicted in the realm of Lynden Bell's
theory are indeed numerically observed, thus confirming the adequacy
of the proposed interpretative scenario.  Note, however, that the
physics is different whether we vary the energy at fixed $f_0$ or at
fixed $M_{0}$. In particular, there is a ``re-entrant'' phase when we
vary the energy at fixed $f_{0}$ \cite{Chavanis_EPJB_06} but there is
no ``re-entrant'' phase when we vary the energy at fixed $M_{0}$
\cite{prl_andreaduccio}.

\subsection{Connection between the two phase diagrams}
\label{connection}

To make the connection between the phase diagram $(f_0,U)$ of
Sec. \ref{fu} and the phase diagram $(M_0,U)$ of Sec. \ref{mu}, we can
plot the iso-$M_{0}$ lines in the $(f_0,U)$ phase diagram or the iso-$f_0$ lines in the $(M_0,U)$ phase diagram.

\begin{figure}
%[htbp]
%  \centering
%\vspace*{3em}
\includegraphics[width=8cm,angle=0]{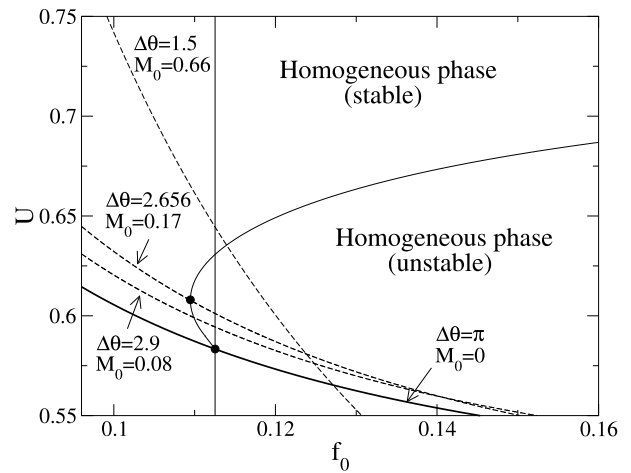}
  \caption{Iso-$M_0$ lines in the $(f_0,U)$ phase
  diagram. This graphical construction allows one to make the
  connection between the $(f_0,U)$ phase diagram of
  Fig.~\ref{muepsilon} and the $(M_0,U)$ phase diagram of
  Fig.~\ref{fig:phase-diagram}. We can vary the energy at fixed
  initial magnetization by following a dashed line. The intersection
  between the dashed line and the curve $U_{min}(f_{0})$ determines
  the minimum energy $U_{min}(M_{0})$ of the homogeneous phase. The
  intersection between the dashed line and the curve $U_{c}(f_{0})$
  determines the energy $U_{c}(M_{0})$ below which the homogeneous
  phase becomes unstable.  } \label{fig:ma}
\end{figure}

Let us first consider the iso-$M_{0}$ lines in the $(f_0,U)$ phase
diagram.  If we fix the initial magnetization $M_{0}$, or equivalently
if we fix the parameter $\Delta\theta$, the relation between the
energy $U$ and $f_0$ is
\begin{eqnarray}
\label{cn1}
U_{\Delta\theta}(f_0)=\frac{1}{6(4\Delta\theta f_0)^{2}}-\frac{1}{2}
\left (\frac{\sin \Delta\theta}{\Delta\theta}\right )^{2}+\frac{1}{2}.
\end{eqnarray}
Therefore, the iso-$M_{0}$ lines are of the form
\begin{eqnarray}
\label{cn2}
U_{\Delta\theta}(f_0)=\frac{A(\Delta\theta)}{f_0^{2}}-B(\Delta\theta),
\end{eqnarray}
with $A(\Delta\theta)=\frac{1}{6(4\Delta\theta)^2}$ and
$B(\Delta\theta)=\frac{1}{2} \left (\frac{\sin
\Delta\theta}{\Delta\theta}\right )^{2}-\frac{1}{2}$, which are easily
represented in the $(f_0,U)$ phase diagram (see
Fig. \ref{fig:ma}). As an immediate consequence of this geometrical
construction, we can recover the minimum energy of the homogeneous
phase for a fixed initial magnetization $M_{0}$ (or
$\Delta\theta$). Indeed, for a given $\Delta\theta$, the homogeneous
phase exists iff $U_{\Delta\theta}(f_0)\ge U_{min}(f_0)$ leading to
\begin{eqnarray}
\label{cn3}
f_0^{2}\le (f_0)_{\Delta\theta}^{2}\equiv \frac{\pi^{2}-(\Delta\theta)^{2}}{48\pi^2\sin^{2}\Delta\theta}.
\end{eqnarray}
This corresponds to $U\ge U_{min}(M_0)=U_{\Delta\theta}((f_0)_{\Delta\theta})$ leading to
\begin{eqnarray}
\label{cn5_1}
U\ge U_{min}(M_0)=\frac{1}{2}\left (\frac{\sin^{2}\Delta\theta}{\pi^{2}-(\Delta\theta)^{2}}+1\right ),
\end{eqnarray}
which is identical to Eq. (\ref{cn5}). Figure \ref{fig:ma} is
in good agreement with the structure of the phase diagram in the
$(M_0,U)$ plane. Indeed, along an iso-$M_0$ line, we find that for
large energies $U>U_{c}(M_0)$ the homogeneous phase is stable and for
low energies $U<U_{c}(M_0)$ the homogeneous phase becomes unstable. In
that case, there is no re-entrant phase. We also note
that for $M_0<(M_0)_*$, the phase transition goes from second order to
first order. This corresponds to the case where the iso-$M_{0}$ line
crosses the tricritical point.

{\it Remark}: we see on Fig. \ref{fig:ma} that
different iso-$M_0$ lines can cross each other. This means that
different initial conditions $(M_0',U)$ and $(M_0'',U)$ can correspond
to the same $(f_0,U)$ hence to the same Lynden-Bell distribution. In
other words, in Lynden-Bell's theory the couples $(M_0',U)$ and
$(M_0'',U)$ are equivalent. There is therefore some redundance in
using the variables $(M_0,U)$ instead of the variables
$(f_0,U)$. Note, however, that the Lynden-Bell prediction does not
always work so that, in case of incomplete relaxation, the couples
$(M_0',U)$ and $(M_0'',U)$ may lead to different QSS.
However, this happens for $f_0 \ge 0.11053\dots$, i.e. in a 
parameter range which is only marginally 
interesting for our analysis.

%%\begin{figure}
%[htbp]
%  \centering
%\vspace*{3em}
%%\includegraphics[width=6cm,angle=-90]{diafasM0U_1.eps}
%%  \caption{Iso-$M_0$ lines in the $(f_0,U)$ phase
%%  diagram. This graphical construction allows one to make the
%%  connection between the $(f_0,U)$ phase diagram of
%%  Fig.~\ref{muepsilon} and the $(M_0,U)$ phase diagram of
%%  Fig.~\ref{fig:phase-diagram}. We can vary the energy at fixed
%%  initial magnetization by following a dashed line. The intersection
%%  between the dashed line and the curve $U_{min}(f_{0})$ determines
%%  the minimum energy $U_{min}(M_{0})$ of the homogeneous phase. The
%%  intersection between the dashed line and the curve $U_{c}(f_{0})$
%%  determines the energy $U_{c}(M_{0})$ below which the homogeneous
%%  phase becomes unstable.  } \label{fig:}
%%\end{figure}

\begin{figure}
   \centering \includegraphics[width = 8cm,angle=-0]{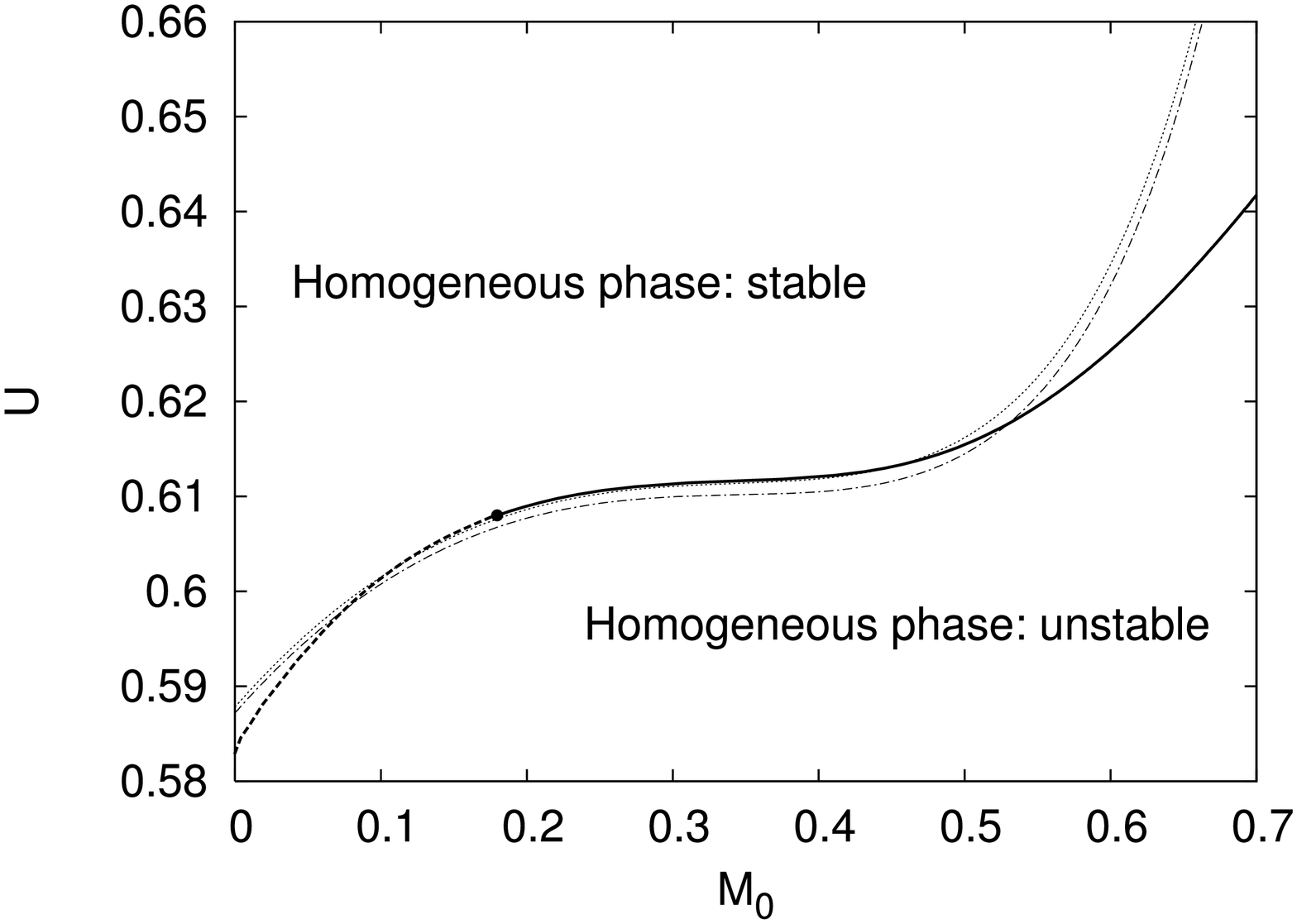}
   \caption{Stability diagram in the $(M_0,U)$ plane, with iso-$f_0$
   lines. The thick lines are the two parts of the stability curve,
   the dotted line is the iso-$f_0$ line with $f_0 =
   0.1096$ and the dash-dotted line is the iso-$f_0$
   line with $f_0 = 0.1100$.}  \label{fig:dfiso}
\end{figure}

Let us now consider the iso-$f_{0}$ lines in the $(M_0,U)$ phase
diagram.  If we fix the phase level $f_{0}$, the relation between the
energy $U$ and $M_0$, or equivalently $\Delta\theta$, is
\begin{eqnarray}
\label{cn1b}
U_{f_0}(\Delta\theta)=\frac{1}{6(4\Delta\theta f_0)^{2}}-\frac{1}{2}
\left (\frac{\sin \Delta\theta}{\Delta\theta}\right )^{2}+\frac{1}{2}.
\end{eqnarray}
Therefore, the iso-$f_{0}$ lines are of the form
\begin{eqnarray}
\label{cn2b}
U_{f_0}(\Delta\theta)=\frac{C(f_0)}{(\Delta\theta)^{2}}-\frac{1}{2} \left (\frac{\sin
\Delta\theta}{\Delta\theta}\right )^{2}+\frac{1}{2},
\end{eqnarray}
with $C(f_0)=\frac{1}{6(4f_0)^2}$, which are easily
represented in the $(M_0,U)$ phase diagram (see Fig. \ref{fig:dfiso}).  Recall that $\Delta\theta$ is
related to $M_{0}$ by
Eq. (\ref{h}). Figure  \ref{fig:dfiso} is
in good agreement with the structure of the phase diagram in the
$(f_0,U)$ plane. In particular, we can see that for a set of values of $f_0$
$\in \lbrack (f_{0})_{*},(f_{0})_{c}\rbrack $, it
intersects the curve $U_c(M_0)$ twice leading to re-entrant phases.
We also note that the  iso-$f_{0}$ lines cannot cross each other contrary to the iso-$M_{0}$ lines.

\subsection{Determination of the absolute minimum energy in the $(f_0,U)$ plane for a rectangular water-bag initial condition}
\label{min}

\begin{figure}
   \centering
      \includegraphics[width = 6cm,angle=0]{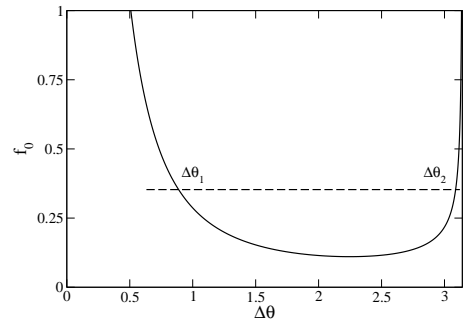}
   \caption{Graphical construction determining the solutions of the equation $U_{f_0}'(\Delta\theta)=0$.}\label{fig:fo}
\end{figure}

\begin{figure}
   \centering \includegraphics[width = 6cm,angle=0]{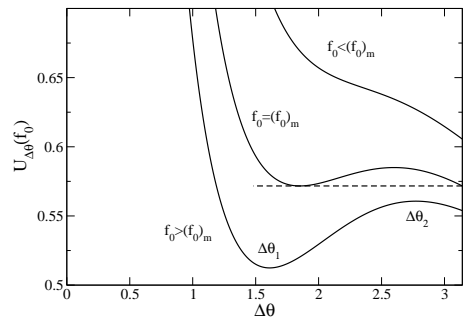}
   \caption{Energy of the initial condition as a function of
   $\Delta\theta$ for different values of $f_0$. From top to bottom:
   $f_0=0.10$, $f=(f_0)_m=0.12135$, $f_0=0.14$. }\label{fig:min}
\end{figure}

We recall that homogeneous Lynden-Bell distributions exist only for
$U>U_{min}(f_0)$. However, there can exist inhomogeneous Lynden-Bell
distributions for $U<U_{min}(f_0)$. Let us determine the minimum
accessible energy $U_{MIN}(f_0)$ when we start from a rectangular
waterbag initial condition. For fixed $f_{0}$ the energy of the initial condition is a function of $\Delta\theta$ (or initial magnetization $M_0$) given by
\begin{eqnarray}
\label{min1}
U_{f_0}(\Delta\theta)=\frac{1}{6(4\Delta\theta f_0)^{2}}-\frac{1}{2}
\left (\frac{\sin \Delta\theta}{\Delta\theta}\right )^{2}+\frac{1}{2}.
\end{eqnarray}
We thus have to determine the minimum of this function for
$0\le\Delta\theta\le \pi$. First of all, the condition
$U_{f_0}'(\Delta\theta)=0$ is equivalent to
\begin{eqnarray}
\label{min2}
f_0=\frac{1}{\sqrt{48\sin(\Delta\theta)\left (\sin(\Delta\theta)-\Delta\theta\cos(\Delta\theta)\right)}}.
\end{eqnarray}
This function is represented in Fig. \ref{fig:fo}. For
$f_0<0.11053...$ there is no solution and for $f_0>0.11053...$ there
are two solutions $\Delta\theta_1$ and $\Delta\theta_2$ corresponding
to one local minimum and one local maximum (see Fig. \ref{fig:min}).  For
$f_0=0.11053...$ , we have $\Delta\theta_1=\Delta\theta_2=2.2467...$ .
Then, we find that the local minimum is the absolute minimum iff
$U_{f_0}(\Delta\theta_1)<U_{f_0}(\pi)$. This is the case if
$f_0>(f_0)_m=0.12135...$ corresponding to an energy $U_m=0.57167...$ (see Fig. \ref{fig:min}). For
$f_0<(f_0)_m$, the absolute minimum correspond to $\Delta\theta=\pi$.

In conclusion, for $f_0<(f_0)_m$, we find that $U_{MIN}(f_0)=U_{f_0}(\pi)$ so that
\begin{eqnarray}
\label{min3}
U_{MIN}(f_0)=U_{min}(f_0)=\frac{1}{96\pi^2 f_0^2}+\frac{1}{2}.
\end{eqnarray}
For  $f_0>(f_0)_m$, we find that $U_{MIN}(f_0)=U_{f_0}(\Delta\theta_{1})$ where $\Delta\theta_{1}$ is the smallest root of Eq. (\ref{min2}). Combining these equations, we find that the absolute minimum energy $U_{MIN}(f_0)$ is given in parametric form by
\begin{eqnarray}
\label{min4}
U_{MIN}=\frac{1}{2}\left (1-\frac{\sin(2\Delta\theta_1)}{2\Delta\theta_1}\right ),
\end{eqnarray}
\begin{eqnarray}
\label{min5}
f_0=\frac{1}{\sqrt{48\sin(\Delta\theta_1)\left (\sin(\Delta\theta_1)-\Delta\theta_1\cos(\Delta\theta_1)\right)}},
\end{eqnarray}
with $0\le \Delta\theta_1 \le 1.85063...$ . For $f_0\rightarrow +\infty$
(non degenerate limit), we get $\Delta\theta_1\rightarrow 0$ and
$U_{MIN}(f_{0})\rightarrow 0$. This is the energy corresponding to an
initial condition $f(\theta,p,t=0)=\delta(p)\delta(\theta)$.

{\it Remark:} for $f_0>0.11053$, we confirm on Fig. \ref{fig:min} that
there can exists several initial conditions with the same $f_0$ and $U$
but a different initial magnetization $M_0$. They lead to the {\it same}
Lynden-Bell prediction.

\section{Numerical results}
\label{sect3}
\begin{figure}
   \centering \includegraphics[width = 6cm,angle=-90]{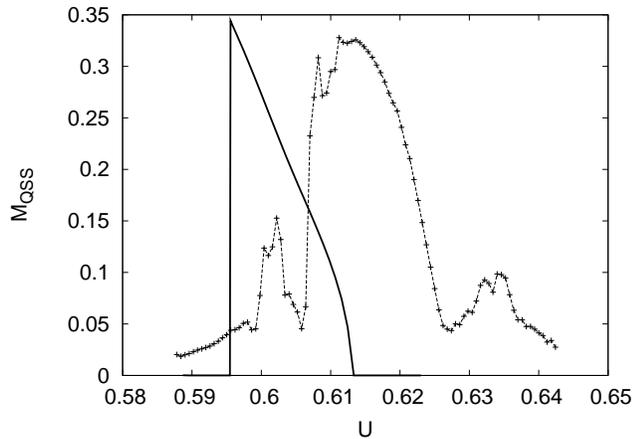}
   \caption{ Magnetization value at QSS, $M_{QSS}$, versus energy for
   $f_0$ = $0.1096$. Comparison between theory (continuous line) and
   simulations (dashed line).  Simulations, done with $N = 10^6$, are
   performed using a symplectic integration algorithm, and averaging
   the magnetization over the time window $40 < t < 140$, and over
   $50$ different realizations. For this $f_0$, $U_{min}=0.5878$,
   $U_r=0.5955$, $U'_c=0.6026$ and
   $U''_c=0.6131$.} \label{fig:confr0}
\end{figure}

To assess the correctness of the above theoretical prediction about the existence of a phase re-entrance, we have performed direct numerical simulations of
the HMF model \eqref{hamil} for finite $N$. For that, we have chosen $f_0$ in the interval $(f_0)_{*}<f_0<(f_0)_{c}$ and ran simulations at
different energies.
Results, for $f_0 = 0.1096$, are shown in Fig. \ref{fig:confr0}, where both the theoretical and numerical values of magnetization at the QSS,
$M_{QSS}$, are plotted as a function of energy.

\begin{figure}
   \centering \includegraphics[width =
   6cm,angle=-90]{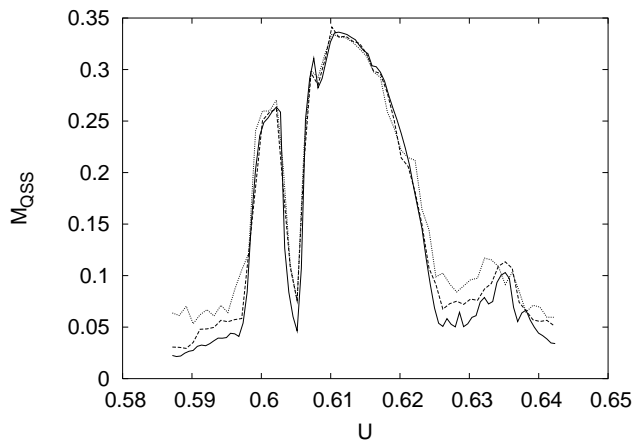} \caption{Magnetization value at
   QSS, $M_{QSS}$, versus energy for $f_0$ = $0.1100$.  For this
   $f_0$, $U_{min} = 0.5872$, $U'_c=0.5980$ and $U''_c= 0.6177$. We
   plotted it for different sizes of the system: the continuous line
   correspond to $N = 10^6$, the dashed to $N =5 \cdot 10^5$ and the
   dotted to $N = 10^5$.} \label{fig:confr2}
\end{figure}
Simulations (dashed line) confirm the existence of a regime of phase
re-entrance. However, the agreement with theory (continuous line) is
mainly qualitative, as there is a systematic shift between the two
curves, although the magnetization value of the main bump is
consistent with the one predicted from Lynden-Bell's
approach. Moreover, simulations show the existence of two zones of
magnetization revival at both sides of the central magnetized
region. If we move at $f_0$ = $0.1100$ (Fig. \ref{fig:confr2}), we
find that one of the two bumps has grown; this confirms that the
structure of the phase diagram is more complex than predicted by the
theory, as we find the existence of additional phase
re-entrances. Simulations performed using different numbers of
particles show that the magnetization values of the central magnetized
region and of the two bumps do not depend on the system size. Instead,
as expected, the curve offset goes to zero when the system size is
increased, see Fig. \ref{fig:confr2}.  

A possible explanation for the discrepancies between theory and
simulations can be found by considering that the energy range in which phase-rentrance is observed is quite narrow. In fact, the iso-$f_0$ lines, in
the interval $(f_0)_{*}<f_0<(f_0)_{c}$, are very close to the
theoretical phase-transition curve, see Fig. \ref{fig:dfiso}. This
means that any possible (i.e., even small) disagreement between the
theoretical and the numerical one may easily lead: a) to a further
numerical phase re-entrance, if the iso-$f_0$ line crosses the
numerical phase-transitions curves without crossing the theoretical
one; b) to a larger numerical value of $M_{QSS}$, if the iso-$f_0$
separates from the numerical curve, while staying close to the
theoretical one. 

\begin{figure}
   \centering
      \includegraphics[width = 6cm,angle=-90]{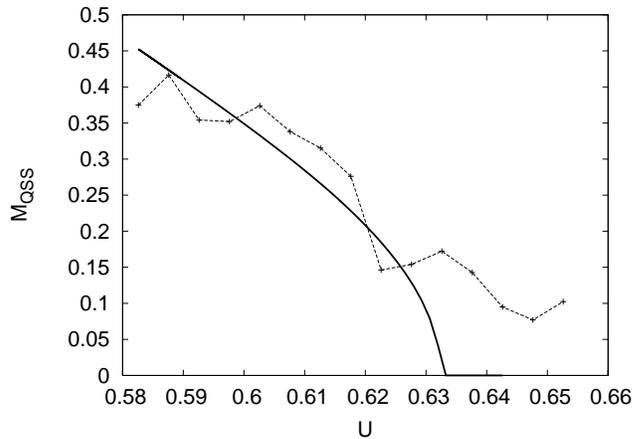}
   \caption{Magnetization value, $M_{QSS}$, versus energy for $f_0=0.1130 > (f_0)_{c}$. Comparison between theory (continuous line) and simulations (dotted line). $U_{min}(f_0 = 0.1130) = 0.5826$ and $U^{\prime \prime}(f_0 = 0.1130) = 0.6325$.}\label{fig:confr}
\end{figure}

\begin{figure}
   \centering \includegraphics[width = 6cm,angle=-90]{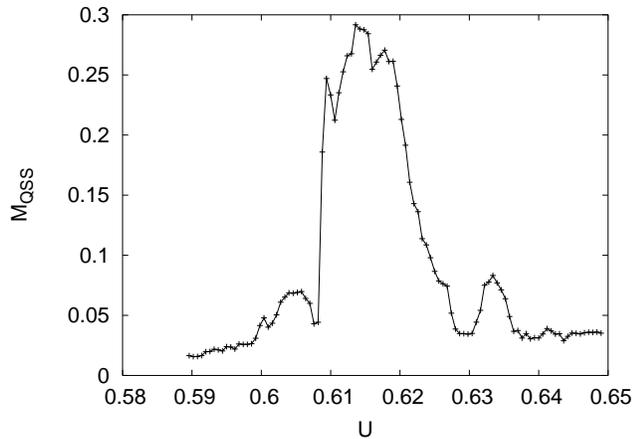}
   \caption{Magnetization value, $M_{QSS}$, versus energy for
   $f_0=0.1085 < (f_0)_{*}$. For $f_0 < (f_0)_{*}$,
   the Lynden-Bell approach predicts that the QSS should be
   non-magnetized so there is a disagreement with theory when $f_{0}$
   is close to the tricritical point.}\label{fig:confr3}
\end{figure}

We also compared theory and simulations for higher $f_0$
($>(f_0)_{c}$). As shown in Fig. \ref{fig:confr}, here theoretical and
numerical results are close. This is in agreement with what is
reported in \cite{prl_andreaduccio}. In Fig. \ref{fig:confr3} we
plotted our numerical results for a $f_0$ lower than
$(f_0)_*$. For $f_{0}=0.1085$, close to the critical
line $U_{c}(f_0)$, we observe a magnetized phase although Lynden-Bell's
approach predicts a non-magnetized phase. For lower values of $f_{0}$,
homogeneous QSS are observed in agreement with the theoretical
prediction (data not shown).

\begin{figure}
   \centering \includegraphics[width = 6cm,angle=-90]{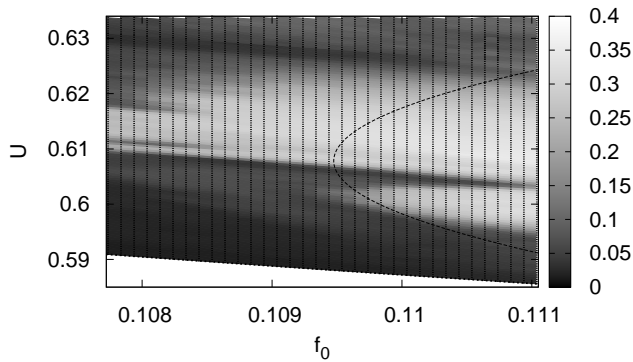}
   \caption{Stability diagram in the $f_0$-$U$ space with numerically
   calculated mean magnetizations. The dashed line is the stability
   curve. The theoretical re-entrant phase is clearly visible as well
   as the second (unexpected) re-entrant phase. In addition to this
   interesting new re-entrance phase, the other main discrepancy is
   the persisting magnetized phase for low $f_0$. } \label{fig:ph_di_col}
\end{figure}

To provide a complete picture of the whole phase diagram, we carried
out simulations on a grid in the $(f_0,U)$ plane and plotted the
numerically obtained values of $M_{QSS}$ in color
scale, see Fig. \ref{fig:ph_di_col}.
%As expected [refs], the agreement between theory and simulations is satisfactory in most of the plane, except in the region $(f_0)_{*}<f_0<(f_0)_{c}$ (see above discussion) and in part of the region with $f_0<(f_0)_{*}$.
At first order, we observe a fair agreement between theory and simulations. In particular, the predicted re-entrant phase phenomenon is clearly observed. This can be considered as a success of the Lynden-Bell theory. We also note that the region (III) of the phase diagram appears to be non-magnetized. It corresponds therefore to a local Lynden-Bell entropy maximum. 
This confirms our claim about the robustness of metastable
states. Note, however, that starting from different initial conditions
(with identical values of $U$ and $f_{0}$), we could have found that
the QSSs in this region are magnetized. Indeed, in the metastability
region, the selection between between local (metastable) or global
(state) entropy maxima depends on a complicated notion of basin of
attraction.
Furthermore, we also find some unpredicted phenomena, as the
additional phase re-entrance, for $U \simeq 0.605$ (see also
Fig. \ref{fig:confr2}) and a persisting magnetized phase for low $f_0$
(also Fig. \ref{fig:confr3}).

\begin{figure}
   \centering
       \includegraphics[width = 5cm,angle=-90] {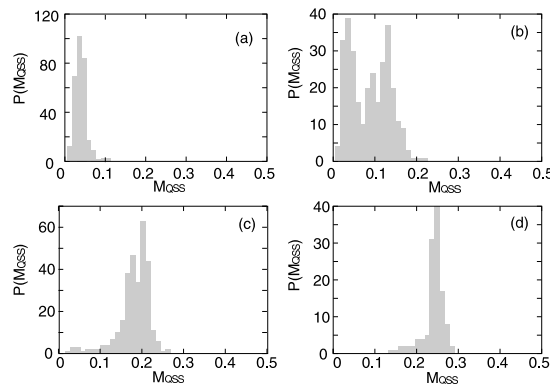}
     \caption{Probability distributions of $M_{QSS}$ for different $U$ values at $f_0=0.1100$. Here in (a) $U= 0.5970$, in (b) $U = 0.5980$,
       (c) $U= 0.5990$, in (d) $U = 0.6000$.}
     \label{fig:multi1}
\end{figure}

\begin{figure}
   \centering
      \includegraphics[width=5cm,angle=-90]{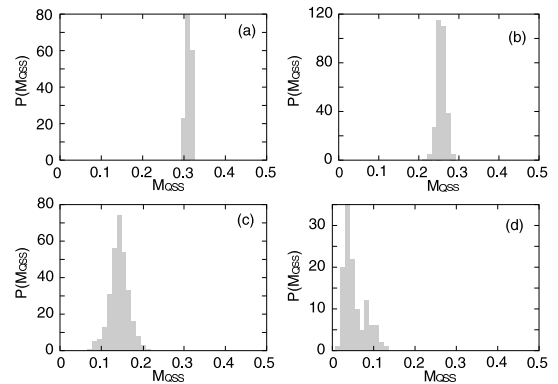}
    \caption{Probability distributions of $M_{QSS}$ for different $U$ values at $f_0=0.1100$.  Here in (a) $U= 0.6150$, in (b) $U = 0.6190$,
       (c) $U= 0.6230$, in (d) $U = 0.6270$.}
     \label{fig:multi2}
\end{figure}

We also studied the order of phase transitions, by plotting the
probability histogram of $M_{QSS}$ sampled with $300$ different
realizations. We show the results for two of the transitions occurring
in Fig. \ref{fig:confr2}. In Fig. \ref{fig:multi1}, one can see that
for the transition at $U \simeq 0.5980$, distributions are
characterized by a double peak, which is a clear signature of a
first-order phase transition. For the one at $U \simeq 0.6230$, see
Fig. \ref{fig:multi2}, the distributions are instead characterized by
a single peak, which validates the prediction of a second-order phase
transition. The two transitions at the boundaries of the smaller phase
re-entrance, occurring around $U \sim 0.605$, not predicted by the
theory, are found to be of first (at low energy) and second (at high
energy) order (data not shown).

\begin{figure}
   \centering
       \includegraphics[width = 6cm,angle=-90] {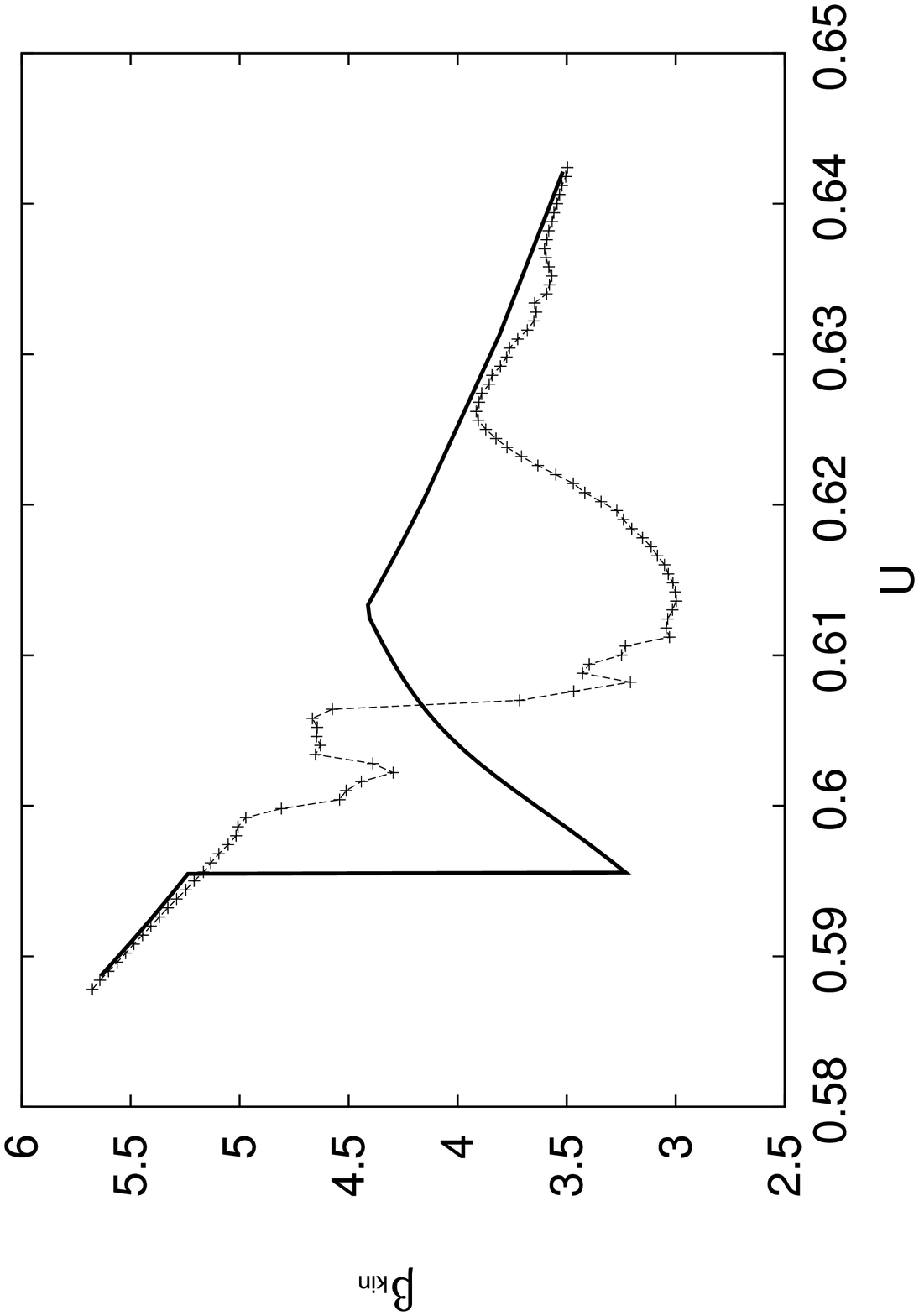}
   \caption{Comparison between theoretical (continuous line) and numerical (dashed line) caloric curves, for  $f_0 = 0.1096$. }\label{fig:cur_cal}
\end{figure}

Finally, we compared the analytical and numerical caloric curves $\beta(U)$ for a given value of $f_{0}$.  In the simulations, the temperature has been calculated from the usual expression   
\begin{equation}
\frac{1}{\beta}_{kin}=< p^2> = \int \mbox{d} \theta \, \mbox{d} p \, \bar{f}_{\text{QSS}}(\theta,p) \, p^2 \label{bl} \;.
\end{equation}
We note that the ``kinetic'' temperature defined by Eq. (\ref{bl})
does not coincide with the Lagrange multiplier $\beta$ associated to
the energy conservation in the Lynden-Bell distribution
(\ref{eq:barf}). This is due to the fermionic nature of this
distribution. Therefore, in order to make the comparison between
simulations and theory relevant, we have calculated the theoretical
temperature from the mean square momentum (\ref{bl}) averaged with the
Lynden-Bell distribution given by
Eq. (\ref{eq:barf}). The results are reported in
Fig. \ref{fig:cur_cal}. In continuity with the results of
Fig. \ref{fig:confr0}, the range of energies where the inhomogeneous
phase appears is shifted with respect to the theoretical
prediction. As a further point, we also notice the
presence of a region with negative specific heat, both in
the numerical and analytical curves. To the best of our
knowledge, this is the first time negative specific heat
is observed out of equilibrium. Surprisingly, this
phenomenon is here observed in correspondence of a second
order transition line.

%In
%this case, the ``kinetic'' temperature does not coicinde with the
%Lagrange multiplier associated to energy conservation, $\beta$. As a
%consequence, the numerically calculated inverse temperature has been
%compared with the theoretical value of the inverse temperature. The
%latter was calculated from the mean square momentum, averaged with the
%Lynden-Bell distribution:
%\begin{equation}
%\frac{1}{\beta}=< p^2> = \int \mbox{d} \theta \, \mbox{d} p \, \bar{f}_{\text{QSS}}(\theta,p) \, p^2 \label{bl} \;,
%\end{equation}
%where $\bar{f}_{\text{QSS}}$ is given by eq. (\ref{eq:barf}). 

\section{Conclusion}
\label{sect4}

In this paper, we have confronted the predictions
\cite{Chavanis_EPJB_06,Antoniazzi_pre_07,prl_andreaduccio,
Antoniazzi_prl_07,reentrant} of a theory based on Lynden-Bell's
statistical mechanics of violent relaxation \cite{Lynden-Bell} to the
results of numerical experiments.  The application of Lynden-Bell's
theory to the HMF model predicts a re-entrant phase in the $(f_0,U)$
plane \cite{Chavanis_EPJB_06} and, indeed, we observe it. It occurs
for a narrow range of parameters which would have been difficult to
find without such a theoretical prediction. In this sense, this is a
great success of Lynden-Bell's approach. The theory also predicts the
correct value of the magnetization in the inhomogeneous phase and the
correct order of the phase transition. This is again remarkable
because the phase diagram displays first and second order phase
transitions in a very narrow range of parameters $(f_0,U)$. All these
predictions are confirmed by direct $N$-body experiments. We have also
numerically observed that metastable states (local Lynden-Bell entropy
maxima) can be very robust, so that they are stable in practice. This
is a specificity of systems with long-range interactions
\cite{Antoni_04, Chavanis_AA_05}.

However, we have also found some discrepancies with respect to
Lynden-Bell's theory. In particular, numerical simulations have
demonstrated the existence of second re-entrant phases: a band of
un-magnetized states in the theoretically magnetized region, as well
as persisting magnetized states in the theoretically un-magnetized
region. As a matter of fact, there is a systematic shift of the
transition line with respect to theory. We must emphasize, however,
the very small selected region of parameters in Figs. \ref{fig:confr0}
and \ref{fig:ph_di_col}.  This gives the impression of a big
discrepancy although the discrepancy is in fact very small.

Therefore, from these numerical experiments, we can conclude that the
Lynden-Bell statistical theory gives a fair first order description of
QSSs in the HMF model. However, for some initial conditions, there can
be more or less severe discrepancies with respect to the
prediction. This is a well-known fact in stellar dynamics
\cite{Lynden-Bell} and vortex dynamics \cite{brands} to which this
theory was initially applied (see a detailed discussion in
\cite{hb3}). Discrepancies with the Lynden-Bell theory have also been
reported for the HMF model in
\cite{Chavanis_EPJB_06} and \cite{bachelard}. These 
discrepancies are usually the result of an {\it incomplete relaxation}
\cite{Lynden-Bell}, i.e. a lack of efficient mixing in the 
system phase space. Indeed, the Lynden-Bell theory is based on a
hypothesis of ergodicity and the prediction fails (by definition) if
the evolution is not ergodic. A detailed understanding of incomplete
violent relaxation is still lacking and appears to be very difficult
\cite{csr}.

Another cause of discrepancy may be related to the proximity of the
numerically considered parameters $(f_{0},U)$ to the critical line and
to the tricritical point. Indeed, it is well-known in equilibrium
statistical mechanics that strong fluctuations are present close to a
critical point, and that the mean field results cease to be valid in
the vicinity of a critical point \cite{kadanoff}. In the present case,
we are studying out-of-equilibrium phase transitions and it is not
clear if we can directly extend equilibrium results to that
situation. Nevertheless, it is not unreasonable to expect that the
theoretical results may be altered close to the critical line and this
is indeed what we observe numerically. Further away from the critical
line (i.e. for larger or smaller values of $f_0$), we find a very good
agreement with the Lynden-Bell prediction (see also
\cite{Antoniazzi_pre_07}). These different observations 
concerning the success
or the failure of the Lynden-Bell theory are consistent 
with the
discussion given in \cite{Chavanis_EPJB_06}.

\end{document}